# La orientación como una seña de identidad cultural: Las iglesias históricas de Lanzarote

# The orientation as a signature of cultural identity: The historic churches of Lanzarote

Alejandro Gangui[1], A. César González García[2], Mª Antonia Perera Betancort[3] y Juan Antonio Belmonte[4]

(1) Instituto de Astronomía y Física del Espacio (CONICET-UBA), Buenos Aires, Argentina.
(2) Instituto de Ciencias del Patrimonio, Incipit, CSIC, Santiago de Compostela, España.
(3) Servicio de Patrimonio Histórico. Excmo. Cabildo Insular de Lanzarote, Canarias, España.
(4) Instituto de Astrofísica de Canarias, La Laguna, Tenerife, España.

Resumen
La orientación de las iglesias cristianas es un elemento distintivo de su arquitectura que repite patrones desde época paleocristiana. Existe una tendencia general a orientar sus ábsides en el rango solar, con una predilección de las orientaciones cercanas al este geográfico (equinoccio astronómico), aunque las alineaciones en sentido opuesto, con el ábside a poniente, si bien resultan excepcionales pues no siguen el patrón canónico, no son inusuales. El caso de las iglesias construidas en el noroeste de África antes de la llegada del Islam resulta paradigmático en este sentido y pudiera reflejar tradiciones anteriores. El Archipiélago canario representa el extremo occidental de esa *koine* cultural norteafricana, por lo que se ha considerado relevante abordar un estudio de un conjunto compacto de iglesias antiguas en alguna de las islas, eligiendo el de Lanzarote. Se ha medido la orientación de un total de 30 iglesias edificadas con anterioridad a 1810, así como algunos ejemplos más de época posterior. La muestra indica que se siguió un patrón de orientación determinante en la isla, pero al contrario que la norma encontrada hasta ahora en el resto del orbe cristiano, este prototipo es doble. Por un lado, aparece la representativa orientación a levante (o poniente), pero la muestra tiene además un patrón marcado de orientaciones hacia el norte-noreste exclusivo, por ahora, de Lanzarote. Se analiza el porqué de esta extraña regla, considerándose varias posibilidades desechadas en su mayoría. Encontramos que la explicación puede ser muy prosaica, de forma que, a veces, las necesidades terrenales resultan más relevantes y decisorias que las necesidades del culto.
Palabras clave: Orientación de iglesias, templos cristianos, Astronomía

Abstract
The orientation of Christian churches is a well-known distinctive feature of their architecture. There is a general tendency to align their apses in the solar range, favoring orientations close to the east (astronomical equinox), although the alignments in the opposite direction, namely, with the apse towards the west, are not unusual. The case of the churches built in northwest Africa before the arrival of Islam is paradigmatic in this regard, and may reflect earlier traditions. The Canary Islands is the western end of this North African cultural *koine,* so we thought it would be relevant to study a compact set of old churches in one of the islands of the archipelago, choosing to start our project with Lanzarote. We have measured the orientation of a total of 30 churches built prior to 1810, as well as a few buildings of later times, nearly a complete sample of all the island Christian sanctuaries. The analysis of this sample indicates that a definite orientation pattern was followed on the island but, unlike the standard one often found in most of the Christian world, this prototype is twofold. On the one hand, the representative orientation to the east (or west) is present. However, the sample has also a marked orientation towards the north-northeast, which is as far as we know a pattern exclusive of Lanzarote. We analyze the reasons for this rule and suggest that one possible explanation could be a rather prosaic one, namely, that sometimes earthly needs are more relevant than religious beliefs.

### 1. Introducción: prolegómenos

El estudio de la disposición de las iglesias cristianas ha interesado desde tiempos pretéritos y recientemente ha cobrado un nuevo auge en la literatura especializada al ser éste un factor representativo de su arquitectura. Según los textos de los escritores y apologetas cristianos tempranos, las iglesias debían situarse siguiendo una determinada orientación, es decir, el sacerdote tenía que situarse mirando hacia el oriente durante los oficios. Así lo reconocen Orígenes, Clemente de Alejandría y Tertuliano y el Concilio de Nicea (325) determinó que así fuera. San Atanasio de Alejandría, también en el siglo IV, expresa que el sacerdote y los participantes deben dirigirse hacia el este, de donde Cristo, el Sol de Justicia, brillará al final de los tiempos (*ecclesiarum situs plerumque talis erat, ut fideles facie altare versa orientem solem, symbolum Christi qui est sol iustitia et lux mundi interentur* […]; para un análisis profundo de las fuentes tempranas y de los métodos de orientación puede verse Vogel (1962).

Sin embargo, estas prescripciones no se muestran del todo claras posibilitando optar entre diversas interpretaciones: ¿se orienta hacia la salida del sol el día que comienza la construcción de la iglesia? ¿O hacia la salida del sol otro día que se considere importante, como puede ser el día del santo patrón de la iglesia? O bien la orientación hacia el este, ¿debe considerarse en sentido estricto? ¿Se orientaban las iglesias hacia la salida del sol en el equinoccio? En ese caso, ¿hacia qué equinoccio?

En un principio, las basílicas cristianas tempranas no se construían con el ábside, o la cabecera de la iglesia hacia el este. A este respecto, Delgado-Gómez (2006) indica que de las 20 primeras basílicas cristianas construidas durante el tiempo de Constantino y sus sucesores en Roma, Jerusalén, Constantinopla y el Norte de África, 18 se sitúan aproximadamente en la línea este-oeste, pero el ábside de 11 de ellas está dirigido hacia el oeste. Sin embargo, es interesante destacar que en estos casos la cátedra y el sacerdote se posicionan contemplando el este, pues el altar está situado entre él y las personas asistentes.

Entre los siglos III y VII se precisan las recomendaciones y así las Constituciones Apostólicas indican que las iglesias se deben construir orientadas hacia el este (*Const. Apost.*, II, 7).[1] En el siglo V, Sidonio Apolinar y Paulino de Nola indican que el ábside debe mirar hacia el este, al equinoccio, algo confirmado más tarde tanto por el Papa Virgilio como por Isidoro de Sevilla en sus *Etymologiae* (XV, 4; McCluskey 1998). Esto sería confirmado durante la Edad Media plena por Honorio Augustodunense (ss. XI-XII, […] *ecclesiae ad orientem vertuntur ubi sol oritur* […]) y por otros autores como Guillermo Durando (ss. XII-XIII, […] *versus orientem, hoc est, versus solis ortum aequinoctialem, nec vero contra aestivale solstitium* […]), que claramente indica la dirección a seguir, el equinoccio, y la que evitar, el solsticio. La orientación hacia el este tiene una clara simbología, como comentamos antes. Es en esa dirección por donde sale el sol y por tanto Cristo, como Sol de Justicia, surgirá desde allí en el Juicio Final (McCluskey 2004, 2010). Por el contrario, la no preferencia de los solsticios podría estar ligada a la importancia de estas fechas en el periodo anterior y a los numerosos templos paganos orientados en esas direcciones (véase, por ejemplo, Belmonte, 2012).

---

[1] Las Constituciones Apostólicas, en su libro segundo, sección séptima, párrafo LVII, señala: "Y sea, primero, el edificio alargado, con su cabecera hacia el este...". Ver
http://www.ccel.org/ccel/schaff/anf07.ix.iii.vii.html (consultado, 14 abril 2015).

Sin embargo, en estas prescripciones, todavía persiste un tanto de ambigüedad al orientar las iglesias hacia el este, pues cabría preguntarse hacia qué equinoccio hacerlo. Como menciona McCluskey (2004) existen varias posibilidades: el equinoccio vernal romano ocurría en el 25 de Marzo, mientras que el griego acontecía el 21 de Marzo –como quedó plasmado en el Concilio de Nicea– ; pero se pueden usar otras definiciones, tales como la entrada del sol en el signo de Aries o el equinoccio de otoño. De cada una de estas definiciones se obtendrían fechas, y por tanto orientaciones, ligeramente diferentes (Ruggles 1999, González-García & Belmonte 2006).

Otro punto importante a considerar es el uso del Calendario Juliano durante la Edad Media y buena parte de la Moderna. La naturaleza de éste haría que, si nos fijamos en un equinoccio calendárico –es decir en una fecha concreta– tal momento se desplazaría con respecto a las estaciones, algo que se vería reflejado en un cambio sistemático de orientación, si ésta se hacía por observación de la salida del sol en ese día.

El estudio de las orientaciones de las iglesias medievales europeas es, junto con las pirámides de Egipto y los megalitos europeos, uno de los ensayos más antiguos que se han trabajado en Arqueoastronomía. González-García (2014) ha llevado a cabo recientemente una recopilación de los trabajos en este campo. En él se aprecia que las prescripciones para la orientación hacia el oriente se siguieron de forma bastante sistemática en toda Europa, al menos durante la Edad Media, como puede verse en la Figura 1. Todas las zonas estudiadas por González-García (2014) siguen ese patrón de orientaciones con un claro máximo predominante centrado en el este, destacando que en numerosas ocasiones, sobre todo en Europa occidental, tal máximo está ligeramente desplazado hacia el norte respecto al este astronómico, tal vez indicando un uso de fechas concretas para el equinoccio (25 de Marzo) que con el paso del tiempo se van trasladando, como se indicó anteriormente, aunque en cada región concurren características particulares.

Resulta interesante destacar que una constante en la literatura sobre la orientación de las iglesias es que éstas se orientan hacia la salida del sol en la efemérides del santo patrón (lo que no parece en absoluto ser el caso de las iglesias lanzaroteñas, como veremos). Sin embargo, en los escritos tempranos y hasta bien entrada la Edad Media no existe un refrendo epigráfico para tal afirmación. Hasta la aparición de las órdenes religiosas en la Baja Edad Media no se constata tal tendencia. Los trabajos revisados por González-García (2014) indican que para áreas de Alemania y tal vez de Inglaterra y Francia podría existir un interés por ciertos santos en algunos monumentos, si bien estos edificios son en general iglesias monacales románicas o catedrales góticas, y por lo tanto tardías. Un caso interesante y bien documentado se da en Eslovenia donde Čaval (2009) ha encontrado evidencias de una predilección especial por la festividad denominada de la Cátedra de San Pedro, inclinación que se ve reflejada en la orientación de un número significativo de iglesias en la dirección del orto solar de ese día. McCluskey (2004) indica que algo similar ocurre en Inglaterra con la orientación de las iglesias románicas donde, tal vez, las iglesias con advocaciones marianas y algunos pocos santos más pueden seguir esta norma de forma complementaria a la orientación hacia el este.

En este contexto, y dado el objeto del presente estudio, resulta interesante destacar que, salvo un número pequeño de trabajos dedicados a iglesias particulares, a sus orientaciones y a posibles eventos de iluminación, sobre todo en Inglaterra y Centroeuropa, no existen estudios sistemáticos sobre la orientación de los templos en períodos posteriores a la Edad Media, como el que nos ocupa, pues como veremos, la gran mayoría de las iglesias y ermitas de Lanzarote se empezó a erigir décadas después de la conquista y colonización de la isla por los normandos al servicio de la corona de Castilla en el siglo XV.

Curiosamente, una excepción a la norma de orientaciones hacia levante es el Norte de África, donde las iglesias se construyeron en direcciones opuestas. Los datos que muestra la Figura 2 fueron obtenidos por Estéban *et al.* (2001) y Belmonte *et al.* (2007), así como otros no publicados con anterioridad (González-García 2014) e incluye un total de 23 iglesias, en particular de África Proconsular y Tripolitania, posibles tierras de origen de la población aborigen canaria (Belmonte *et al.* 2010). Es interesante observar que se constata un buen número de iglesias con orientaciones hacia poniente, costumbre usual en los momentos tempranos del cristianismo, como se señaló anteriormente. También se destaca que la mayoría de las iglesias se ordenan dentro del rango solar, con concentraciones en los equinoccios y los solsticios, lo que podría dar claves sobre el proceso de cristianización en esta región.

En España, tanto en la Península Ibérica como en los dos archipiélagos, si bien existen informaciones de eventos particulares de iluminación dentro de templos románicos en momentos especiales como el equinoccio (como en Santa Marta de Tera o en San Juan de Ortega, en las provincias respectivas de Zamora y Burgos), la cuestión de la orientación de las iglesias ha sido poco investigada en general, desde un punto de vista estadístico, lo que ha llevado a afirmaciones de tipo un tanto peregrino respecto a la posible causa de las desviaciones de algunas iglesias con respecto a la orientación canónica (véase por ejemplo Godoy-Fernández, 2004).

Pérez-Valcárcel (1998) ha investigado la orientación de 187 iglesias románicas del Camino de Santiago. Aunque sus datos no incluyen la medida de la altura angular del horizonte, algo desgraciadamente muy común en otros estudios europeos, lo que sí parece claro es que no se establece una relación general entre la orientación de estas iglesias y la salida del sol en la fecha del santo patrón de advocación de la iglesia.

Nuestro equipo ha decidido iniciar un proyecto a gran escala tanto en la península como en el Archipiélago Canario. En este último, este trabajo es el primer estudio sistemático desarrollado. Sin embargo, dentro de un programa más amplio para medir las orientaciones de las iglesias prerrománicas del territorio peninsular de forma sistemática, González-García *et al.* (2013) han dedicado una especial atención a las iglesias del periodo asturiano y a su interacción con el poder musulmán dominante en el sur de la Península. En concreto, encuentran que las 13 iglesias del periodo aún existentes en Asturias poseen una orientación canónica, con el ábside hacia oriente, aunque desviada en general varios grados al norte del este. Además, los autores han encontrado que las mezquitas de Al-Ándalus, si bien podrían haberse orientado hacia La Meca, con quiblas que podrían haber sido compatibles con las alineaciones canónicas de las iglesias, siguen sin embargo otras disposiciones. Por un lado, un buen número de mezquitas se orientan hacia el sudeste, mientras que otras siguen la de la Mezquita de Córdoba, ambas disposiciones permitidas por el Islam. Así, parecería que las mezquitas "evitan" orientaciones posibles que puedan confundir sus templos con iglesias, mientras que las iglesias asturianas, y tal vez las mozárabes inmediatamente posteriores también, tenderían a evitar aquellas posiciones que confundan estos templos con mezquitas, en un ejemplo de la interacción de religión, poder y astronomía. Por tanto, vemos que ante circunstancias excepcionales, los patrones canónicos pueden ser alterados.

Por último, García-Quintela *et al*. (2013) han investigado la introducción del cristianismo en el noroeste de la Península y la posible sustitución de elementos culturales indoeuropeos (célticos) por factores cristianos, mediante la introducción de lo que denominan un "paisaje mártir": a través de las orientaciones de las iglesias y la cristianización de sus entornos, así como de la creación de mitos y relatos que canalizan, modifican o sustituyen a los posibles cultos paganos. Sería pues interesante analizar esta misma fenomenología en el archipiélago canario y, en particular, en la isla de Lanzarote, un caso especialmente llamativo dado que fue la primera en ser colonizada por Europa y su tamaño y número de núcleos poblacionales permitiría un estudio de una muestra estadísticamente significativa en un espacio muy compacto y reducido.

## 2. Las iglesias y ermitas de Lanzarote

La arquitectura religiosa en la isla de Lanzarote comenzó con la edificación de ermitas de factura sencilla y de recinto único, como la iglesia–catedral de San Marcial incluida en la *Zona Arqueológica San Marcial de Rubicón* y de la que se conserva su pavimento interior, cimientos y arranque de sus paredes, en el sector de Papagayo en el sur insular. A algunas de ellas, con el paso del tiempo, se les fueron agregando capillas en la cabecera, sacristías a sus lados y otros elementos de uso práctico como barbacanas y calvarios (ver Figura 3). En general, estas construcciones no se sujetan a planes de ejecución estrictos; así, tanto su planta como su estructura se levantaron de acuerdo a las necesidades del momento. Con el tiempo, algunas alcanzaron un cierto carácter monumental.

Generalmente responden a la siguiente organización: portada en arco de medio punto a los pies, remarcada en piedra del lugar, aunque a veces se encuentra una segunda portada en un costado; espadaña de uno o varios huecos; techo a dos o cuatro aguas enlucido o con tejas; paramento exteriormente enlucido de blanco y barbacana, en ocasiones sencilla otras más prominente, ubicada en alguno de sus extremos, presumiblemente para cortar el viento. En su interior, las ermitas pueden poseer su nave y altar a diferente altura, y cubrirse con sencillos techos de madera de estilo mudéjar, en artesa o a cuatro faldones, a lo máximo seis, frecuentemente con lacerías.

Dado el elevado número de monumentos históricos, cercano a la treintena y por tanto estadísticamente significativo, se eligió Lanzarote como primer laboratorio de prueba donde estudiar la orientación de las iglesias canarias en los siglos inmediatamente posteriores a la conquista y analizar si en esta ubicación habían influido factores como la presencia de la población aborigen en las islas (Jiménez González 2009, 2011), con unos patrones de culto y sistemas de cómputo de tiempo totalmente diferentes a los de las personas recién llegadas (Belmonte *et al*. 1994).

La Tabla 1 muestra los datos obtenidos en una campaña de trabajo de campo en junio de 2012. Se presentan las cantidades estándar (identificación de la iglesia o ermita y sus coordenadas) junto con las referencias de su orientación (datos arqueoastronómicos): acimut y altura angular del horizonte (medidos), y la declinación correspondiente (calculada). En lo referente a las fechas consignadas para las iglesias, existe cierta ambigüedad: para algunas se dispone de datos precisos de edificación, en otros casos el año es estimado, aproximado o bien corresponde a la primera mención documentada del edificio, mientras que en otras las fechas no han podido ser precisadas. Sin embargo, estas dataciones pueden sernos útiles para el análisis posterior de los datos.

Obtuvimos las mediciones empleando un par de tándems que incorporan un clinómetro y una brújula de alta precisión, analizando asimismo el entorno paisajístico de cada uno de los edificios. Posteriormente corregimos los datos de acuerdo a la declinación magnética local. Nuestros valores de la declinación magnética para distintos sitios de la isla oscilan desde 4°38' a 4°46' oeste. Como la precisión que tenemos con los acimuts magnéticos medidos es de ¼° aproximadamente, hemos tomado para los 32 valores una misma declinación magnética, igual a 4°42' oeste. Los valores establecidos para los acimuts y alturas angulares son el promedio de dos y a veces hasta tres mediciones y debemos enfatizar que, con pocas excepciones, las diferentes mediciones diferían en menos de ½°. En cualquier caso, dadas las alteraciones magnéticas consignadas en varios lugares de la isla (sobre todo en el entorno de las erupciones volcánicas de Timanfaya), se han verificado algunas medidas con imágenes fotosatelitales, encontrándose pocas divergencias. Por ello, estimamos que el error de nuestras medidas está en torno a los ±¾° como cota superior, siendo así los datos lo suficientemente adecuados para culminar el estudio estadístico deseado.

En la Figura 4 se muestra el diagrama de orientación para las iglesias y ermitas. Como hemos mencionado, los valores de los acimuts son los medidos, e incluyen la corrección por declinación magnética. Las líneas diagonales del gráfico señalan los acimuts correspondientes −en el cuadrante oriental− a los valores extremos para el sol (acimuts de 62°,5 y 116°,7 −líneas continuas−, equivalente a los solsticios de verano e invierno, respectivamente) y para la luna (acimuts: 56°,4 y 123°,8 −líneas rayadas−, equivalente a la posición de los lunasticios mayores).

De las 32 ermitas e iglesias medidas, 12 se encuentran orientadas en el cuadrante norte, 2 en el cuadrante a poniente, 17 orientadas en el cuadrante a levante (13 de ellas en el rango solar) y tan solo 1 en el cuadrante meridional (ver Figura 4). Aunque sabemos que la muestra no es representativa de todas las islas del Archipiélago Canario (aunque sí lo es de la isla de Lanzarote), aquí se distinguen dos orientaciones claras: (i) hacia el norte (con "entrada" a sotavento, evitando quizás los vientos dominantes del lugar), y (ii) hacia levante, con el ábside de la ermita apuntando hacia el cuadrante oriental. Las 13 orientaciones *ad orientem* entran dentro de la lógica observada en otros estudios sobre orientaciones de iglesias, pero en éste resulta notable la cantidad de orientaciones hacia el cuadrante norte y que caen fuera del rango solar. Parece tratarse de un caso absolutamente singular de la isla de Lanzarote donde lo práctico y prosaico (los alisios provenientes del norte-noreste, Figura 5) parece combinarse (fuertemente) con lo cultual y canónico (rango solar).

Respecto al rango solar, o más bien lunisolar, destacan dos particularidades. Por un lado tenemos la iglesia matriz de la capital histórica, Teguise (ver Figura 6) orientada prácticamente en la dirección del orto más meridional posible de la luna (una orientación ya encontrada en el mundo aborigen en casos tan paradigmáticos como el Roque Bentaiga o Montaña Tindaya; Belmonte y Hoskin 2002). Por otro, se destaca la iglesia de Ntra. Señora de las Mercedes en Mala (Fig. 3), única ermita de la isla orientada con precisión al equinoccio (la de Los Dolores parece equinoccial pero con mayor discrepancia) y que además parece seguir una norma infrecuente en la isla, la de estar orientada al orto solar en el día de su advocación (Tabla 1). En este sentido, también se destaca que la mayoría de las iglesias de Teguise (ver, por ejemplo, Fig. 3), la antigua capital de la isla y donde se establecieron la mayoría de los colonizadores, se orientan (con la notable excepción de la iglesia matriz de Guadalupe, de orientación "anómala" según se ha dicho) a declinaciones incluidas en el rango canónico. Si bien el equinoccio también era importante en el mundo aborigen (Belmonte *et al.* 1995), parece ser que en el lugar elegido por la mayoría de la población europea para establecerse se respetaron las normas provenientes de sus lugares de origen.

Para intentar comprender mejor lo expuesto, en la Figura 7 presentamos el histograma de declinaciones, que resulta independiente de la ubicación geográfica y de la topografía local. Éste muestra la declinación astronómica frente a la frecuencia normalizada, lo que permite una más clara y acertada determinación de la estructura de picos, marcando su relevancia dentro del mismo. De nuevo, el pico asociado a esas orientaciones hacia el norte-noreste, absolutamente excepcional, domina el gráfico. Cuando nos encontramos con este fenómeno en los datos, nos planteamos cuál pudo haber sido la causa, estableciéndose diferentes interpretaciones: ¿pudo ser el resultado de una influencia aborigen? (Belmonte y Hoskin, 2002) ¿Tuvo algo que ver el poblamiento de la isla con la esclavitud morisca importada de la cercana costa africana y de tradición islámica? (Anaya Hernández 2004, 2008). Sin embargo, sería la explicación más prosaica la que nos daría la clave. Para comprender esto tenemos que hacer un análisis más detallado de cierta característica geográfica de la isla.

### 3. El Jable, una historia de desolación

El Jable es un amplio sector del centro de la isla, caracterizado por poseer un suelo cubierto de una capa sedimentaria de variable grosor, compuesta por arenas orgánicas provenientes de los fondos marinos que, una vez alcanzada la playa, son arrastradas por los vientos dominantes de componente norte–sur (ver Figura 8). Estas arenas acceden a la isla por Bahía de Penedo y Playa de Famara y la traspasan hasta alcanzar su costa centro oriental. La consecuencia es una franja de suelo dinámico de 5 km de ancho que atraviesa Lanzarote de norte a sur en su parte central, a modo de pasillo cubierto de estas arenas móviles frágiles a los efectos del viento si en ella no crece vegetación que contribuya a inmovilizarla. Su contextura geológica va variando al entremezclarse con arenas basálticas principalmente, e igualmente su grosor cambia y algunos sectores poseen una considerable potencia, convirtiéndose en suelos aptos para la producción agraria o ganadera principalmente ya que permite el crecimiento de una cobertura vegetal que constituye el alimento del ganado compuesto generalmente de cabras y en menor medida ovejas. Los límites territoriales de este ecosistema han experimentado una persistente modificación a lo largo de los siglos, siendo esencial para ello la presencia de un tapiz vegetal que impida o dificulte la movilidad de la arena, como ya hemos adelantado. Del mismo modo, este ecosistema arenoso en actividad constituye un peligro para la población, al sepultar cultivos, aldeas y viviendas diseminadas e incluso a lo largo de los siglos ha propiciado la alteración del emplazamiento de diversas localidades como Muñique, Mozaga o San Bartolomé. A pesar de ello, un conjunto de núcleos poblacionales a lo largo de la historia se asienta en sus límites o en su interior para favorecer su aprovechamiento.

Distintas fuentes documentales e históricas (ver, por ejemplo, León Hernández y Robayna Fernández 1989, León Hernández, Robayna Fernández y Perera Betancort 1990) nos informan sobre este peculiar ecosistema; extensión, deslinde de propiedades cuyas divisorias han resultado sepultadas, tala de arbustos que aceleran el avance de las arenas perjudicando de esta manera los cultivos y los pastos, tormentas de jable que sotierran aldeas, casas, tierras de cultivo, obligando a su reubicación, etc.

En la primera visita pastoral del obispo Cristóbal de la Cámara y Murga (entre 1627 y 1629) se recoge que "[...] *En todas he estado, sin faltar uno, ni Iglesia, o Ermita que no haya visitado, visto y tocado todo por mis ojos y manos [...]* [2] detallando que en Lanzarote […] *hay unos grandísimos montones de arena, que va corriendo entre dos mares insensiblemente que ven caminar de manera que como las aguas son, que salen de mar y vuelven a entrar en él, y es tan grande la altura de la arena, que se podrían hundir a partes seis hombres, de allí corren a menos de legua de la Villa* […]. Interesa observar que este clérigo secular de origen vasco visita todos los templos de la isla edificados en ese entonces que son la ermita de Nuestra Señora del Socorro (se construye en el año 1612), San Juan Bautista en Haría (edificado en 1625),[3] y San Andrés en la localidad de Tao, término municipal de Teguise (citada en el año 1627 y caída en 1735) sin advertir anomalía alguna referida a su orientación, puesto que su sínodo persigue la finalidad de […] *corregir las costumbres y establecer el régimen espiritual de la Iglesia, conforme al espíritu del Concilio tridentino* […].[4]

En 1770 en el documento de la Real Audiencia (Leg. 10.749s/f) del Archivo Histórico Provincial de Las Palmas se lee que en El Jable […] h*ay una sierra de arena a manera de río que va dominando aquel término e impide la producción de pasto. Si no se impide la producción de terrazgos se quedarán sin recursos de pastos ni leña y por consiguiente se despoblarán aquellos lugares [...]. Este tal jable de arena sobre distan más de una legua de tierras cultivadas por componerse el término* [El Yagabo] *de más de 2.000 fanegas siendo menos de 200 las que se panifican no impide dicha arena la producción de los matos que se aprovechan en leña y en orden a los acueductos de las maretas* […]. Del mismo modo en el Compendio Anónimo de 1776[5] se refleja que Lanzarote: *Hallase dividida naturalmente en dos partes, por que reduzido su ancho (quassi a la mitad de ella) a solas quatro leguas por una ensenada que haze el mar en la parte del norte, en donde llaman boca de Famara, por aquí corre y desde aquí atrabesando asi al sur una riacha de arenas blancas que agitada de los vientos hazen continuo del su curso y formas una faxa que asi por su figura blanca y angosta como por su única situación a la mitad de la Ysla en el lugar más humilde y havatido de ella parece quisso naturaleza señalar esta de su cuerpo la delicada cintura, y el penado de ambas partes que desde este citio se empezaban a elevar y a formar por sus medianías una reprochosa aunque ancha cumbre de grandes colinas o montañas que se habrasan y unen con siertas sierras* [...].

El día 12 de mayo de 1818 el personero Juan Valenciano Curbelo[6] atiende a esta zona a la que define como de: […] *arena blanca menuda que aquí llamamos jable, que agitada por los repetidos vientos, corre como un río, y atravesando de norte a sur la isla entera extiende cada día sus ruinosos límites* […].

---

[2] Constituciones Sinodales del Obispo de Canarias. 1629.
[3] Ignoramos la alineación del primer templo de Nuestra Señora de la Encarnación fechado aproximadamente en el año 1561 y derribado en 1618.
[4] Ídem.
[5] Anónimo, con introducción y notas de Francisco Caballero Mújica, Ayuntamiento de Teguise. 1991: 15.
[6] Archivo privado de Juan Antonio Martín Cabrera, Las Palmas de Gran Canaria. Petición del personero Juan Valenciano Curbelo a S.M., 12 de mayo 1818. Citado por Perera Betancort (2004).

Del mismo modo, las Actas del Ayuntamiento de Teguise[7], término municipal al que mayoritariamente pertenece El Jable, recogen los efectos naturales de este ecosistema de suelo movedizo. Así podemos leer en los Acuerdos del Ayuntamiento de la Villa de Teguise: […] *y los vientos imperiosos haciendo correr velozmente las arenas que arroja el mar por la parte del norte de la isla y corren hasta la del sur en esta dirección inutilizó muchos terrenos y los extendió por otros que jamás podrán volverse a cultivar, de suerte que las ruinas perdidas y desgracias que ha sufrido ahora esta pequeña isla unida a la que experimentó el año de treinta de siglo pasado con los "espantosos" bolcanes que corrieron en ella por el espacio de siete años continuos, cubriendo la tercera parte de su superficie y con los que bolbieron a repetirse en el año anterior de ochocientos veinte y cuatro la han reducido a la última miseria, por cuyo motivo y a causa de los años calamitosos que han experimentado en este siglo se ven obligados imperiosamente sus naturales a emigrar a las Américas para livertarse de las desgracias que les amenaza* […]. Pocos años más tarde, en 1834, en este mismo Libro Capitular del Ayuntamiento de Teguise,[8] se recoge el deslinde de las arenas […] *Y con respecto a otro particular que comprende lo prevenido por el Señor Gobernador sobre el acortamiento de las Dehesas que pertenecen a este Pueblo sobre lo cual mandó que desde luego dicte el Ayuntamiento las mas enérgicas providencias para impedir que rocen los arbustos que se crían en ella y se contenderá un Reglamento de las medidas que convengan adoptar para conservar las acotaciones penas en las que incurren los contraventores y tarifas de lo que habran de pagar los ganaderos por el pasto en el caso de que no puedan arrendarse. Y convencido el ayuntamiento mas y mas de la necesidad y utilidad que se digne de llevarse a efecto con energía y firmeza estas disposiciones por el convencimiento que tiene y es público y notorio a esta Isla de los grandes daños y ruinas que causan a los Pueblos y propiedades el movimiento, corriente extención de las arenas, se principie desde luego el expediente y el deslinde, amojonamiento y acotamiento que ejecutó el Sr. Presidente al que acompañarán los caballeros diputados* [...].

Baltasar Perdomo, cura de San Bartolomé en 1830 dibuja una interesante cartografía de El Jable que recoge Eduardo Hernández Pacheco[9] […] *La mancha blanca que atraviesa la isla de Norte a Sur son las arenas del Jable que han inutilizado casi todos estos terrenos que eras feraces y algunos de los mejores de la isla, como la Vega de Mosaga, Regla, Bebederos, etc. Los terrenos que ocupan las dos líneas amarillas descendiendo de la playa de la Caleta a la playa Honda, eran los límites de las arenas hasta el año de 1800, que desde aquella época se han ido extendiendo así a una y otra parte de los parajes colindantes: 1) el paraje donde fue el lugar de Mosaga y hoy solo queda su ermita y un vecino y los demás se han pasado sobre el volcán; 2) casas ya arruinadas por dichas arenas, así en San Bartolomé como en Corral de Guirrez; 3) casas de los señores Terrens, Carrasco, González y Tejera, donde ya tocan las arenas; 4) campos que se hallan cubiertos de arbustos, los que impedían extenderse estas arenas, que arrojaban y arrojan las playas de la Caleta y Famara y han causado los estragos que se ven en los campos limítrofes por haberlos desmontado; 5) montañas de arenas movedizas que llamamos médanos, 6) dónde deben hacerse paredes de dirección. [Hecho por el cura de San Bartolomé el día 3 de diciembre de 1830, en la isla de Lanzarote]*.

---

[7] Archivo Histórico de Teguise.1826: 25
[8] Archivo Histórico de Teguise.
[9] Eduardo Hernández Pacheco. 1909: 299 – 301.

En estos documentos trasciende la recurrencia a la tala de arbustos y matos que crecen espontáneamente e incluso se arranca la barrilla (*Messenbryantemun nodiflorum*) para su exportación destinada a la fabricación de jabón. Un reflejo de este hecho se halla en los Acuerdos del Ayuntamiento de Teguise[10]: [...] *Algunos magnates* [...] *intentaron el secuestro de los cuatro términos denominados Soo, Bajamar, Muñique y Cuchillo o Caldera de Juan Pérez* [en el año 1824], [...] *a la codicia de la cosecha de la barrilla* [...]. *Que se deje crecer las aulagas* [*Launaea arborens*] *y se prohíba el desmonte de leña y se ponga un guarda* [...] *si no se desmonta, la aulaga que evita correr el jable* [...]. Este hábitat, de interés para Leonardo Torriani y José Viera y Clavijo, se ha destinado al cultivo al menos desde el siglo XVII, como así lo refleja el escribano Figueras en 1620 mencionando a Fiquinineo, lugar emplazado en la parte central de El Jable. No obstante ya el propio término vega lleva implícito la actividad agrícola de este suelo: [...] *Juan Cabrera, vecino de Lanzarote, vende a Baltasar Rodríguez, vecino de esta isla, en la aldea de Tiagua; conviene a saber de una suerte de tierra para labranza que yo tengo en la vega de Fiquinineo, de buen dote de casamiento con Dñª. María Betancor, mi mujer.* [...][11].

Por su parte Leonardo Torriani[12] al redactar sobre la isla puntualiza: *Esta isla no tiene árboles, pero está llena de matorrales que dicen tabaibas. Del Norte hacia el Sur, empezando desde Famara, la atraviesan montículos de arena, los cuales del mismo modo que las arenas líbicas son llevados por el viento septentrional.* [...]. En su Diccionario de 1845–50, Pascual Madoz[13] se refiere a *Jable (El): porción despoblada de la isla de Lanzarote* [...] *dicha proporción del terreno inundado* [...] *El terreno inundado será como de 5 leguas cuadradas que en la actualidad nada producen, aunque la mayor parte fuese productivo antes, como la vega de Mozaga, la de Soo y la de Fiquinineo. En el término de Soo y Bajamar se hallaban detenidos este inmenso depósito de jable hace algunos siglos por los matos* [...]. Este mismo autor[14] recoge: *Soo: vega en la isla de Lanzarote, situada al W. del término de Bajamar y al S. de la montaña y pueblo de Soo. Consta de unas 3.000 fanegas de tierra –inundadas de jable–, de las que solo unas 200 contiguas a aquél producen si hay un poco de invierno abundante centeno y barrilla.* [...].

Mencionemos, por último, a René Verneau, quien visita la isla hacia 1884[15] y revela que acude a la zona por recomendación de Baltasar Perdomo, cura de San Bartolomé que ya se ha mencionado: *Me recomendó vivamente que fuera a explorar una aldea recientemente destruida por un huracán que, después de haber derrumbado las casas, la recubrió de arena. Fiquinineo [así se llamaba] fue un pueblo habitado por las sacerdotisas de Venus.* [...].

Por tanto, y luego de todos estos testimonios, se destaca el papel fundamental que en la vida de la isla desempeñó durante siglos la existencia del Jable.

**4. Orientación de iglesias y paisaje**

Retomemos ahora el tema que vinimos desarrollando en las secciones previas, el de las alineaciones. Entendemos que esta particularidad en las orientaciones de las iglesias lanzaroteñas tiene poco correlato con otros estudios previos ya mencionados en la introducción. En la Península Ibérica y en todo el Mediterráneo los rangos son predominantemente solares. En especial, la gran proporción de iglesias orientadas aproximadamente hacia el norte resulta novedoso; es el caso, por ejemplo, de la iglesia Nuestra Señora de las Nieves, en La Montaña, patrona de la isla, o el de otras muchas de factura antigua, como las de Tiagua (ver Figura 9) o Tao. Es destacable que una notable proporción de iglesias orientadas de esta manera se hallan en el sector noroccidental y central de la isla, como puede apreciarse en la Figura 10.

---

[10] Archivo Histórico de Teguise 1834: 150.
[11] Escribano Figueras, J. de, Legajo 2722. Archivo Histórico Provincial de Las Palmas, año 1620: 160.
[12] Leonardo Torriani: 1978: 46.
[13] Pascual Madoz. 1846: 128.
[14] Ídem. Pág. 192.
[15] René Verneau. 1981: 126.

La diferencia entre los resultados que arroja el estudio en Lanzarote y otros ya culminados, nos obliga a buscar explicaciones alternativas para intentar comprender el patrón de orientaciones de las iglesias de esta isla. Si éstas en general no se orientan con el sol, ¿podría deberse a una causa tan prosaica como la necesidad de orientar el porche de las construcciones en dirección contraria a los fuertes vientos del noreste dominantes en la isla? (ver Figs. 3 y 5) o, en caso contrario, protegerlo (ver Figs. 3 y 9). O bien, ¿podría deberse a la topografía (quizás cambiante con el tiempo) de diferentes regiones de la isla? En todo caso, parece claro, observando los gráficos y las imágenes, que la cuestión ambiental es relevante.

En lo concerniente a los vientos, las zonas donde se han construido mayor número de iglesias orientadas al norte-noreste (con la puerta a mediodía) es al borde del Jable (noroeste y centro de la isla), donde se vuelve imperioso evitar la arena arrastrada por el viento, a veces en furiosas tormentas como la de 1824 que sepultó varios caseríos, y que aún hoy en día, a pesar de las alteraciones del paisaje, hace notar sus efectos (Perera Betancort 2004 y 2009; Perera Betancort *et al.* 2004). Curiosamente, el mayor número de orientaciones canónicas (es decir, a levante) se dan en edificios situados en el noreste de la isla, al socaire de los vientos, en zonas protegidas de la arena por los riscos de Famara.

En lo relacionado con la topografía cambiante de la isla, las erupciones volcánicas de Timanfaya, ocurridas entre 1730 y 1736, quizás también hayan podido jugar un rol, como veremos más adelante. Por ejemplo, la ermita de Santa Catalina, en Los Valles, responde a una construcción del s. XVIII con una orientación muy peculiar, que sustituyó a otra anterior dedicada al mismo culto, que había sido destruida durante las lavas de Timanfaya. La ermita actual se orienta con acimut 339½° es decir en dirección norte-noroeste, pero está protegida por las montañas que la circundan por lo que es anómala bajo cualquier punto de vista. Desde luego, sería interesante saber la orientación de la iglesia primigenia. En cualquier caso, estas ideas parecen atractivas, pues sugieren que las necesidades humanas a veces pueden superar e imponerse a las obligaciones del culto, lo que resulta excepcional.

Para comprobar alguna evolución notoria en las características de las construcciones durante un largo periodo de tiempo, se ha llevado a cabo un estudio sobre la evolución de la orientación con respecto a la edad. En los distintos paneles de la Figura 11 se representan los valores del acimut y de la declinación en función de las fechas probables de construcción de las iglesias y ermitas, consignados en la Tabla 1. El panel (a) incluye todas las iglesias más antiguas de las que conocemos estas fechas (28 de las 32 de la muestra). El panel (b) resalta aquellas iglesias orientadas dentro del rango solar, ubicadas entre las dos líneas horizontales (de acimuts 62°,5 y 116°,7).

El tercer panel (c) muestra los valores de la declinación en función de la fecha (con líneas horizontales de declinación a −23°,5 y +23°,5). Se destaca la construcción de iglesias en el rango canónico a lo largo de todo el periodo; aunque a partir del segundo cuarto del siglo XVII se empieza la construcción a gran escala de ermitas orientadas hacia el norte, quizás en un momento en que las necesidades humanas superaban con mucho a las de culto al haberse ya familiarizado con el medioambiente isleño. En el gráfico también destacan las erupciones de Timanfaya pues la construcción de ermitas casi cesa en la primera mitad del siglo XVIII, reactivándose poco después, pero de nuevo mostrando el doble patrón.

## 5. Conclusión

Tras la conquista y colonización de la isla canaria de Lanzarote por parte de la población europea a principios del siglo XV, se comenzó en los siglos inmediatamente posteriores la colonización a gran escala de la isla con el establecimiento de pequeñas haciendas y caseríos, junto a algunas villas mayores como Teguise o Femés, donde se inició la construcción de un número no desdeñable de templos cristianos que ilustraban la nueva situación social y religiosa.

En algunos pocos lugares, es posible que se orientasen los edificios con patrones de imitación del culto aborigen. En otros, se respetó la tradición canónica de alinear los templos a levante (con algunas excepciones a poniente) aunque con un grado de tolerancia mucho mayor que el habitual. En este sentido, se debe destacar que solo una iglesia de Lanzarote, la de Mala, parece presentar una orientación compatible con el orto solar en el día de la advocación (mariana) del templo (Fig. 3).

Finalmente, en Lanzarote hay un número estadísticamente significativo de iglesias orientadas en dirección norte-noreste, lo que resulta una notoria excepción a la regla. Se han analizado diferentes posibilidades para explicar esta anomalía, llegándose a la conclusión de que la respuesta más plausible es a su vez la más prosaica. Este patrón de orientación parece obedecer al deseo de evitar los fuertes vientos dominantes en la isla, procedentes precisamente de esa dirección, y, en particular, soslayar las molestias causadas por la arena desplazada por el viento en aquellas edificaciones más cercanas o limítrofes con El Jable.

Este es solo el primer experimento de un proyecto que esperamos poder acometer en los próximos años, midiendo la orientación de los templos cristianos más antiguos en otras islas del Archipiélago Canario. En este sentido, suponemos que será sumamente interesante el estudio de la isla de Fuerteventura, sometida al mismo flujo de viento, incluso más intenso, que el que sopla en la isla vecina de Lanzarote. ¿Tendrán las iglesias de Fuerteventura también un doble patrón? ¿Se atrevieron sus constructores a incumplir el precepto canónico para imponer las necesidades humanas a las del culto? En vista de lo hallado hasta ahora, nos atrevemos a aventurar que sí.

## 6. Bibliografía

Anónimo (1991). *Compendio brebe y fasmosso, histórico y político, en que* [se] *contiene la situación, población, división, gobierno, produziones, fabricas y comercio que tiene la Ysla de Lanzarote en el año de 1776*, con introducción y notas de Francisco Caballero Mújica, Ayuntamiento de Teguise.

Anaya Hernández, L. A. (2004). "El Corso Magrebí y Canarias. El último ataque berberisco a las islas: la incursión a Lanzarote de 1749". *Actas X Jornadas de Estudios sobre Lanzarote y Fuerteventura.* Servicio de Publicaciones del Excmo. Cabildo Insular de Lanzarote. Litografía Romero. Santa Cruz de Tenerife. Tomo I. Págs. 13–29.

Anaya Hernández, L.A. (2008). "La liberación de cautivos de Lanzarote y Fuerteventura por las Órdenes Redentoras. *Actas XII Jornadas de Estudios sobre Lanzarote y Fuerteventura.* Servicio de Publicaciones del Excmo. Cabildo Insular de Lanzarote. Vol. I. Tomo I. Litografía Romero. Santa Cruz de Tenerife. Págs. 65 – 93.

Belmonte J.A. (2012) *Pirámides, templos y estrellas: astronomía y arqueología en el Egipto antiguo*, Crítica, Barcelona.

Belmonte, J.A y Hoskin, M. (2002). *Reflejo del Cosmos: atlas de arqueoastronomía en el Mediterráneo occidental*, Equipo Sirius, Madrid.

Belmonte, J.A., Estéban, C., Aparicio, A., Tejera Gaspar, A., Gónzalez, O. (1994) "Canarian Astronomy before the conquest: the pre-hispanic calendar", *Rev. Acad. Can. Ciencias*. VI, 2-3-4, 133-156.

Belmonte J.A., Estéban C., Schlueter R., Perera Betancort M.A., González O. (1995). "Marcadores equinocciales en la prehistoria de Canarias". *Noticias del IAC*, 4-1995, 8-12.

Belmonte J.A., Tejera A., Perera M.A. y Marrero R. (2007) "On the orientation of pre-Islamic temples of North-west Africa: a reaprisal. New data in Africa Proconsularis", *Mediterranean Archaeology and Archaeometry* 6, 3: 77-85.

Belmonte J.A., Perera Betancort M.A. y González García A.C. (2010) "Análisis estadístico y estudio genético de la escritura líbico-bereber de Canarias y el norte de África", en *VII Congreso de patrimonio histórico: inscripciones rupestres y poblamiento del Archipiélago Canario*, Cabildo de Lanzarote, Arrecife, en prensa.

Cámara y Murga C. de la. *Constituciones Sinodales del obispo de Gran Canaria, y su Santa Iglesia con su primera fundación y traslación, vida sumaria de sus obispos y breve relación de todas las siete islas. 1629.* Imprenta Viuda de Juan González. Madrid 1634. Archivo de la Catedral de Canarias. Las Palmas de Gran Canaria: Archivo Secreto.

Čaval, S. (2009). "Astronomical orientations of Sacred Architecture during the Medieval period in Slovenia", en J.A. Rubiño-Martín, J.A. Belmonte, F. Prada and A. Alberdi (eds.), *Cosmology Across Cultures*, 209-19. San Francisco. Astronomical Society of the Pacific.

Delgado-Gómez, J. (2006). "El porqué de la orientación de las iglesias", *Lucensia* 16, 33. 347-56.

Estéban, C., Belmonte, J.A., Perera Betancort, M.A., Marrero, R. y Jiménez González, J.J. (2001). "Orientations of pre-Islamic temples in North-West Africa", *Archaeoastronomy* 26, S65-84.

García Quintela, M.V., González-García, A.C. y Seoane-Veiga, Y. (2013). "De los solsticios en los castros a los santos cristianos: la creación de un paisaje mártir en Galicia", *Madrider Mittelungen*, in press.

González-García, A.C. (2014). "A voyage of christian medieval astronomy: symbolic, ritual and political orientation of churches", en F. Pimenta, N. Ribeiro, F. Silva, N. Campion, A. Joaquinito, L. Tirapicos (eds.): *Stars and stones*. British Archaeology reports, in press.

González-García, A.C. y Belmonte, J.A. (2006). "Which Equinox?" *Archaeoastronomy, The Journal of Astronomy in Culture* 20. 97-107.

González-García, A.C., Belmonte J.A. and Costa-Ferrer, L. (2013). "The orientation of pre-Romanesque churches in Spain: Asturias, a case of power re-affirmation", en M.A. Rappenglueck, B. Rappenglueck and N. Campion (eds.), *Astronomy and Power*. British Archaeology Reports, in press.

Godoy Fernández, C. (2004). "A los pies del templo. Espacios litúrgicos en contraposición al altar: una revisión", *Antigüedad Cristiana* 21, 473-89.

Hernández Pacheco, E. (1909). "Estudio geológico de Lanzarote y de las isletas canarias". *Memoria de la Real Sociedad Española de Historia Natural.* T. VI.

Jiménez González, J.J. (2009). "Un modelo arqueológico adaptativo en Lanzarote y Fuerteventura a través de sus manifestaciones rupestres", en Actas de las XIV Jornadas de Estudios sobre Lanzarote y Fuerteventura. Arrecife (Lanzarote), 21-25 de septiembre de 2009, en prensa.

Jiménez González, J.J. (2011). "Manifestaciones rupestres y organización tribal en Fuerteventura y Lanzarote", en Actas de las XV Jornadas de Estudios sobre Fuerteventura y Lanzarote. Puerto del Rosario (Fuerteventura), 19-23 de septiembre de 2011, en prensa.

León Hernández, J. de y Robayna Fernández, M.A. (1989). "El Jable, poblamiento y aprovechamiento en el mundo de los antiguos mahos de Lanzarote y Fuerteventura. *Actas de las III Jornadas de Estudios sobre Fuerteventura y Lanzarote.* Servicio de Publicaciones del Excmo. Cabildo Insular de Fuerteventura. Tomo II. Págs. 11 – 105.

León Hernández, J. de, Robayna Fernández, M.A., y Perera Betancort, M.A. (1990). "Aspectos arqueológicos y etnográficos de la comarca del Jable. *Actas de las II Jornadas de Historia de Lanzarote y Fuerteventura.* Servicio de Publicaciones del Excmo. Cabildo Insular de Lanzarote. Tomo II. Págs. 283 – 319.

Madoz, P. (1846). *Diccionario geográfico estadístico histórico de España y sus posesiones de ultramar.* Tomo IV. Madrid.

McCluskey, S.C. (1998). *Astronomies and cultures in early Medieval Europe*. Cambridge University Press. Cambridge.

McCluskey, S.C. (2004). "Astronomy, Time, and Churches in the Early Middle Ages", in M.-T. Zenner, Villard's legacy: *Studies in Medieval Technology, Science and Art in Memory of Jean Gimpel*. Ashgate, Aldeshot: 197-210.

McCluskey, S.C. (2010). "Calendric cycles, the eighth day of the World and the orientation of English Churches", en C. Ruggles and G. Urton (eds.), *Skywatching in the Ancient World, New Perspectives in Cultural Astronomy*, 331-353. University Press of Colorado. Bolder.

Perera Betancort, F.M. (2004). *"Aportación al problema de El Jable a principios del siglo XIX"*. *Actas, X Jornadas de Estudios sobre Lanzarote y Fuerteventura.* Servicio de Publicaciones del Excmo. Cabildo Insular de Lanzarote. Litografía Romero. S.L. Santa Cruz de Tenerife. Tomo I. Págs. 205 – 212.

Perera Betancort, F.M. (2009). *Arquitectura tradicional y elementos asociados de Lanzarote. Asociación para el Desarrollo Rural de la isla de Lanzarote.* Nueva Gráfica, S.A.L.

Perera Betancort, M.A., Marrero Romero, R. y García Navarro, M. (2004). "El yacimiento arqueológico de Ajey. Intervención arqueológica. Fase I. Lanzarote". *Actas X Jornadas de Estudios sobre Lanzarote y Fuerteventura.* Servicio de Publicaciones del Excmo. Cabildo Insular de Lanzarote. Litografía Romero. Santa Cruz de Tenerife. Tomo I. Págs. 487 – 510.

Pérez-Valcárcel, J. (1998). "La orientación de las iglesias románicas del Camino de Santiago", en F. Bores, J. Fernández, S. Huerta, E. Rabasa, *Actas del Segundo Congreso Nacional de Historia de la Construcción*. La Coruña, Servicio de Publicaciones Universidad de La Coruña: 391-396.

Ruggles, C.L.N. (1999). "Whose equinox?" *Archaeoastronomy* 22:S45-50.

Torriani, L. *Descripción e historia del reino de las islas canarias, antes Afortunadas, con el parecer de sus fortificaciones.* Ed. Goya. Santa Cruz de Tenerife, 1978.

Verneau, R. (1981). *Cinco años de estancia en las islas Canarias.* Capítulo III, edición J.A.D.L. La Orotava. Santa Cruz de Tenerife.

Vogel, C. (1962). "Sol aequinoctialis. Problemes et tecnique de l'orientation dans le culture chretien". *Revue Sciences Religieuses* 36, 175-211.

Viera y Clavijo, J. de (1982). *Noticias de la Historia General de las Islas Canarias* [1772 – 1783]. Introducc. y notas de A. Cioranescu, Santa Cruz de Tenerife. Volumen I y II.

VV.AA. (1998). *Patrimonio de Canarias: Lanzarote y Fuerteventura*, Gobierno de Canarias, Las Palmas.

TABLA 1. Orientaciones de las ermitas e iglesias de Lanzarote, obtenidas en nuestra misión de junio de 2012. Para cada construcción, la tabla muestra la ubicación, la identificación (nombre y fecha más probable de construcción), la latitud y longitud geográficas (L y l), el acimut astronómico (a) tomado a lo largo del eje del edificio en dirección al ábside (redondeado al ½° de error), la altura angular del horizonte (h) en esa dirección (0 B significa horizonte bloqueado, se toma h=0°), y la declinación resultante correspondiente (δ).

| Ubicación | Nombre / fecha | L (°/') Norte | l (°/') Oeste | a (°) | h (°) | δ (°) | Fecha santo / orientación |
|---|---|---|---|---|---|---|---|
| (1) Femés | San Marcial de Limoges (1630) | 28/54 | 13/46 | 52½ | 1¼ | 32¾ | 16 Abril - 8 Julio / ---- |
| (2) Yaiza | Ntra. Sra. de los Remedios (1699) | 28/57 | 13/46 | 77½ | 2¾ | 12 | vv.ff. 9 Sept / 21 Abr – 22 Ago |
| (3) Uga | San Isidro (1956) | 28/57 | 13/44 | 15 | 3½ | 60 | 15 Mayo / ----- |
| (4) Masdache | La Magdalena (s. XX) | 28/59 | 13/39 | 339½ | -½ | 54¼ | 22 Julio/ ----- |
| (5) La Geria | Ntra. Sra. de la Caridad (1706) | 28/58 | 13/43 | 44½ | 0½ | 38¾ | 8 Sept R 15 Agos / ----- |
| (6) Mancha Blanca | Ntra. Sra. de los Dolores (1782) | 29/02 | 13/41 | 92 | 1¾ | -1¼ | 15 Sept / 17 Mar – 26 Sept |
| (7) Tinajo | San Roque (1669) | 29/04 | 13/40 | 34½ | -¾ | 45¼ | 16 Agos R/ ----- |
| (8) Yuco | Ntra. Sra. de Regla (1663) | 29/03 | 13/39 | 356½ | -1¼ | 59 | vv.ff. / ----- |
| (9) Tiagua | Ntra. Sra. Del Socorro (1625) | 29/03 | 13/38 | 12 | -1¼ | 57¾ | 8 Sept / ----- |
| (10) Sóo | San Juan Evangelista (1749) | 29/05 | 13/37 | 142 | 0¾ | -43½ | 27 Dic / ----- |
| (11) Tao | San Andrés (1627) | 29/02 | 13/37 | 6½ | -1 | 58¾ | 30 Nov / ----- |
| (12) Mozaga | Ntra. Sra. de la Peña (1785) | 29/01 | 13/36 | 278 | 1¼ | 7½ | R 8-13 Agos / 8 Abr – 4 Sept |
| (13) San Bartolomé | San Bartolomé (1661) | 29/00 | 13/36 | 26½ | 0 B | 51 | 24 Agos R / ----- |
| (14) Nazaret | Ntra. Sra. de Nazaret (1648) | 29/02 | 13/34 | 105 | 1¾ | -12¼ | R 26 Ago / 17 Feb – 25 Oct |
| (15) Teguise | San Rafael (1661) | 29/04 | 13/34 | 72½ | 1¾ | 15¾ | 29 Sept / 4 May – 9 Ago |
| (16) Teguise | El Cristo de la Vera Cruz (1625) | 29/04 | 13/33 | 82 | 0 B | 6¾ | 3 Mayo / 7 Abr – 5 Sept |
| (17) Teguise | San Juan de Dios y San Fco. de Paula (Sto. Domingo) (1698) | 29/03 | 13/34 | 254½ | 0½ | -13¾ | 8 Mar – 2 Abr / 11 Feb – 1 Nov |
| (18) Teguise | Ntra. Sra. de Guadalupe (1680) | 29/04 | 13/34 | 128½ | 5½ | -30½ | 6 Sept / ----- |
| (19) Teguise | Ntra. Sra. Miraflores, Convento de San Francisco (1588) | 29/03 | 13/33 | 84 | 6½ | 8 | ? – 4 Oct / 10 Abr – 2 Sept |
| (20) Teseguite | San Leandro (1674) | 29/03 | 13/32 | 71 | 0¾ | 16½ | 13 Nov / 6 May – 6 Ago |
| (21) El Mojón | San Sebastián (1661) | 29/04 | 13/31 | 42½ | 0¾ | 40¼ | 20 Ene / ----- |
| (22) Los Valles | Santa Catalina (1749) | 29/05 | 13/31 | 339½ | 14¼ | 65¼ | 25 Nov – 20 Abr / ---- |
| (23) La Montaña (Teguise) | Ntra. Sra. de las Nieves (1661) | 29/06 | 13/32 | 15½ | -1 | 56 | 5 Agos / ----- |
| (24) Haría | San Juan (1625) | 29/09 | 13/30 | 98½ | -½ | -7¾ | 24 Jun / 1 Mar – 12 Oct |
| (25) Haría* | Ntra. Sra. de Encarnación (1631) | 29/09 | 13/30 | 74 | 4½ | 15¾ | 25 Mar / 4 May – 9 Ago |
| (26) Mala | Ntra. Sra. de las Mercedes (1809) | 29/06 | 13/28 | 89½ | -½ | 0 | 24 Sept / 20 Mar – 23 Sep Equinoccio |
| (27) Guatiza | El Cristo de las Aguas (1915) | 29/04 | 13/29 | 107½ | 1 | -15 | R 13 Sept / 8 Feb – 3 Nov |
| (28) Arrecife | San Ginés (1570) | 28/57 | 13/33 | 53½ | 0 B | 31¼ | R 25 Agos / ---- |
| (29) Tahiche | Santiago Apóstol (1779) | 29/01 | 13/33 | 70 | 12½ | 23 | 25 Jul / 11 Jun – 2 Jul |
| (30) Tías | Ntra. Sra. de la Candelaria (1795) | 28/58 | 13/39 | 323½ | 8¾ | 49¾ | 2 Feb / ----- |
| (31) Conil | María Magdalena (1794) | 28/59 | 13/40 | 118½ | 12 | -18½ | 22 Jul / 27 Ene – 15 Nov |
| (32) Tegoyo | Sagrado Corazón de Jesús (1863) | 28/58 | 13/41 | 52½ | 9¼ | 36¾ | Móvil, Junio / ---- |

(*) Derribada y reconstruida en el s. XX, presumiblemente con la misma orientación

**FIGURAS Y PIES DE FIGURA**

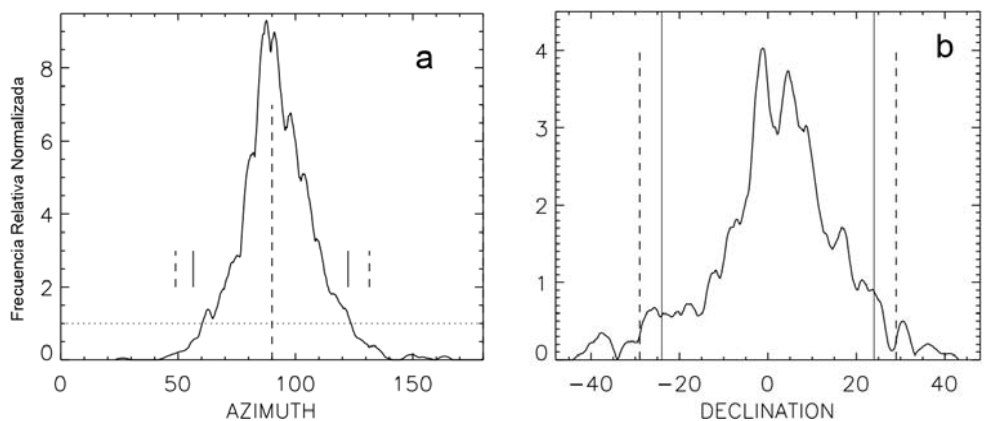

*Figura 1.* Histogramas de acimut (***a***) y declinación (***b***) de orientación de una muestra muy significativa de iglesias medievales europeas. El diagrama de acimut incluye las medidas de 1274 iglesias obtenidas de la literatura, mientras que el de declinación incluye 425 iglesias. Nótese la concentración de orientaciones hacia el este, aunque ligeramente desplazadas respecto al equinoccio astronómico ($\delta=0°$).

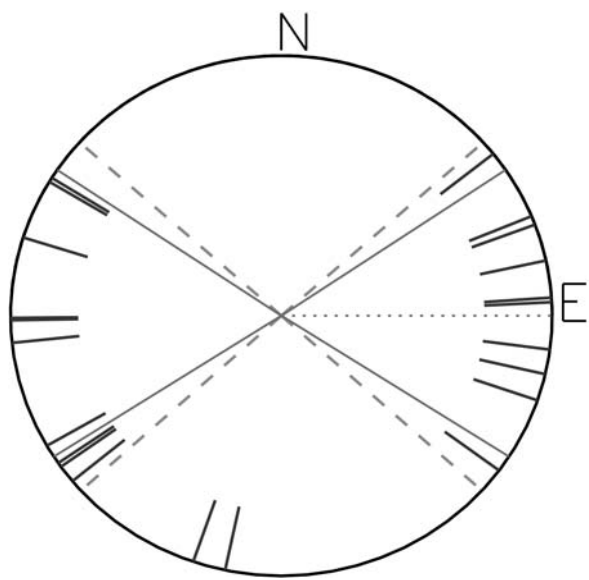

*Figura 2.* Diagrama de orientación del ábside de 23 iglesias tempranas del Norte de África. Salvo dos localizadas en Sbeitla que siguen el patrón ortogonal de la ciudad romana, todas las demás se orientan dentro del rango lunisolar, aunque el rango occidental es mucho más frecuente de lo habitual.

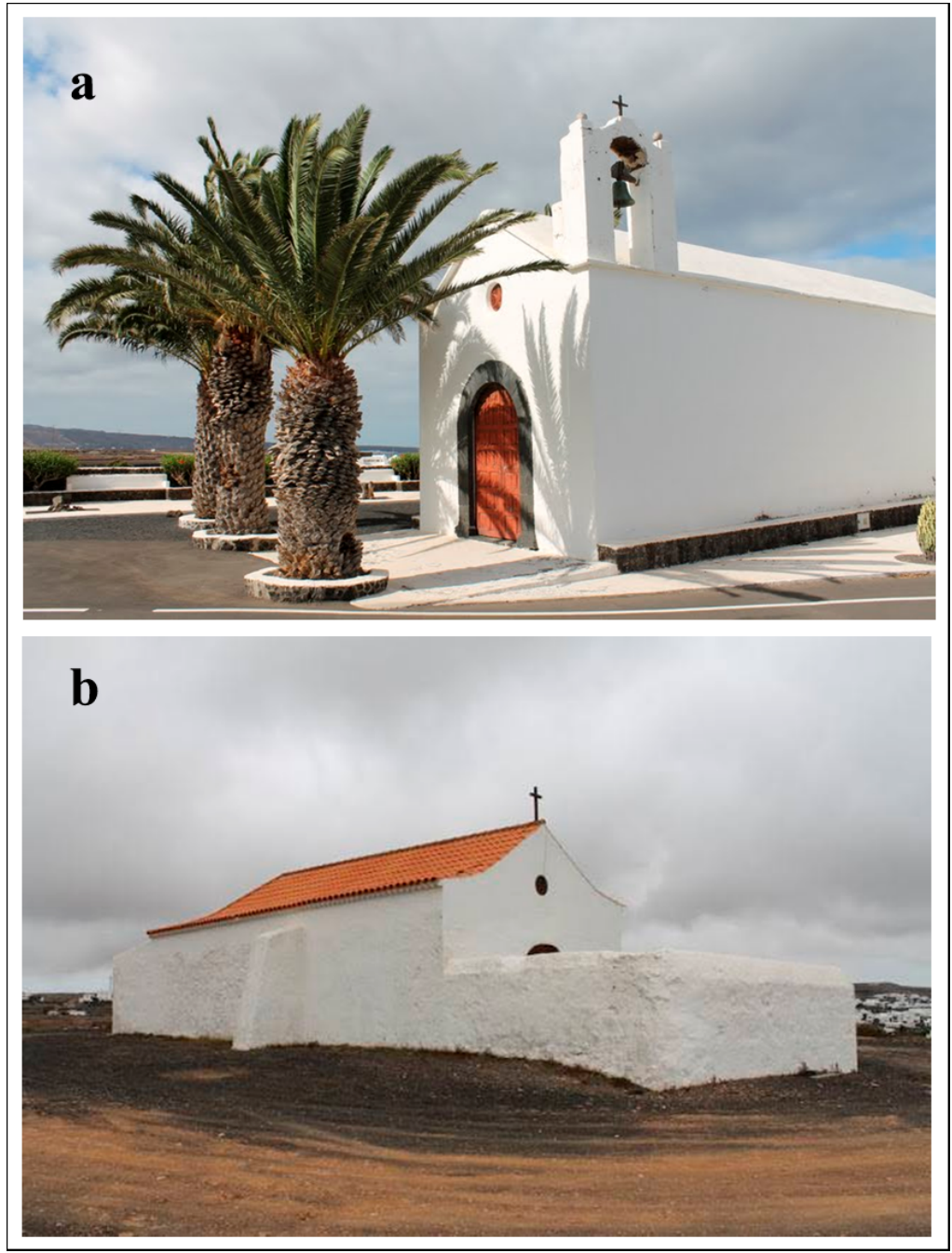

***Figura 3.*** *Dos iglesias de Lanzarote con características singulares: (a) la iglesia de Ntra. Señora de las Mercedes en Mala es la única iglesia de la isla orientada equinoccialmente de forma precisa y que además podría estar orientada al orto solar el día de su advocación mariana (24 de septiembre), una tradición aparentemente extraña al entorno isleño; (b) la ermita de San Rafael, situada sola y aislada sobre una meseta que domina El Jable en las afueras del núcleo urbano de la villa de Teguise; data del s. XIX pero aparece ya citada en 1661. Orientada a levante, su peculiaridad es que posee una gran barbacana en forma de "L" que protege la entrada de los vientos dominantes. Nótese la arena depositada sobre ésta. Imágenes de Alejandro Gangui.*

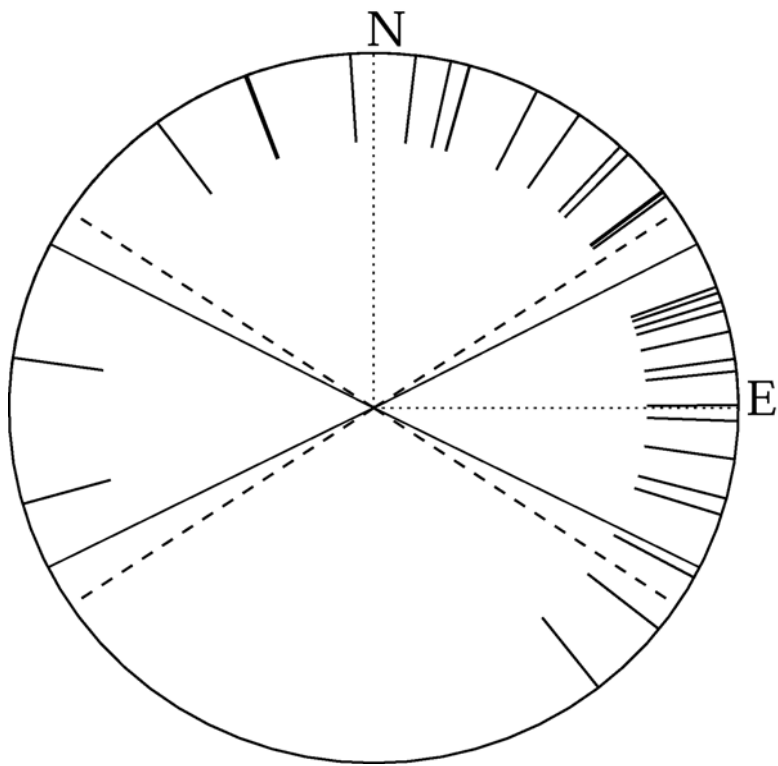

*Figura 4. Diagrama de orientación para las iglesias y ermitas de Lanzarote, obtenido a partir de los datos de la Tabla. 1. Aunque un número significativo sigue el patrón canónico de orientación en el rango solar, un número nada desdeñable de iglesias están orientadas hacia el norte-noreste. Este patrón es exclusivo de Lanzarote y requiere una explicación.*

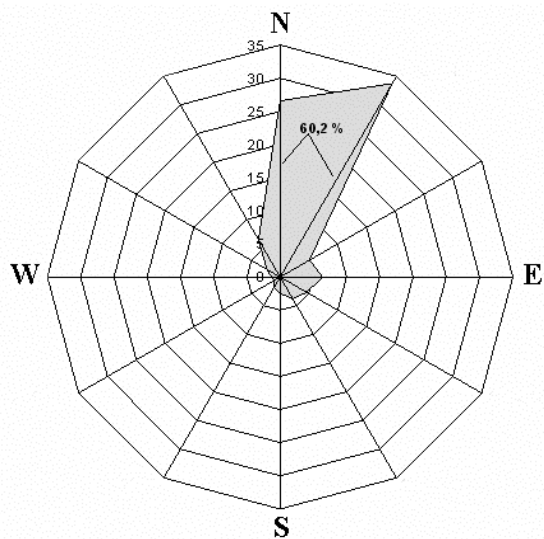

*Figura 5. Diagrama de vientos para el Aeropuerto de Arrecife de Lanzarote, ilustrativo de los vientos dominantes en la isla. Se puede apreciar la enorme concentración en el rango N-NE (acimuts entre 345° y 45°) similar a las orientaciones excepcionales de varias iglesias de la isla. Imagen de archivo.*

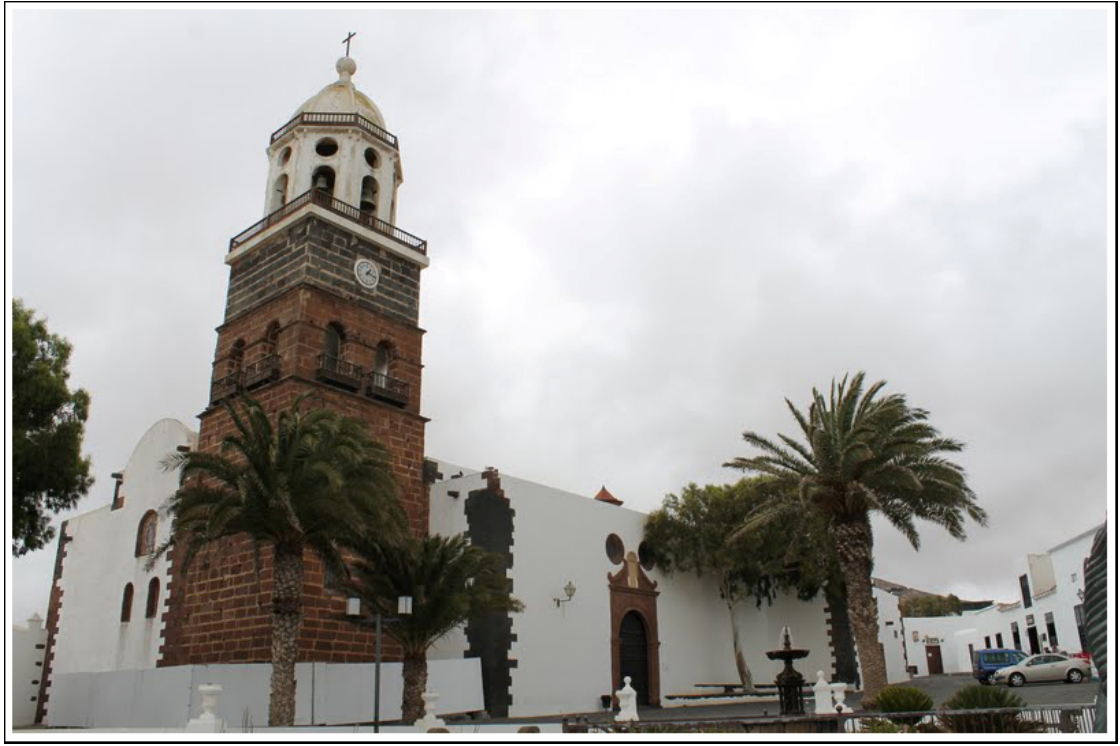

***Figura 6.*** *Una de las iglesias más antiguas e importantes de la isla de Lanzarote es la de Ntra. Señora de Guadalupe en Teguise, cuya fábrica actual data de 1680 pero que se erige sobre templos anteriores. Su orientación no es canónica al estar orientada prácticamente a la posición más meridional del orto lunar, una tradición ajena a las iglesias europeas, pero que quizás refleje tradiciones anteriores teniendo en cuenta que Teguise era la capital de la isla ya en tiempos preeuropeos. Imagen de Alejandro Gangui.*

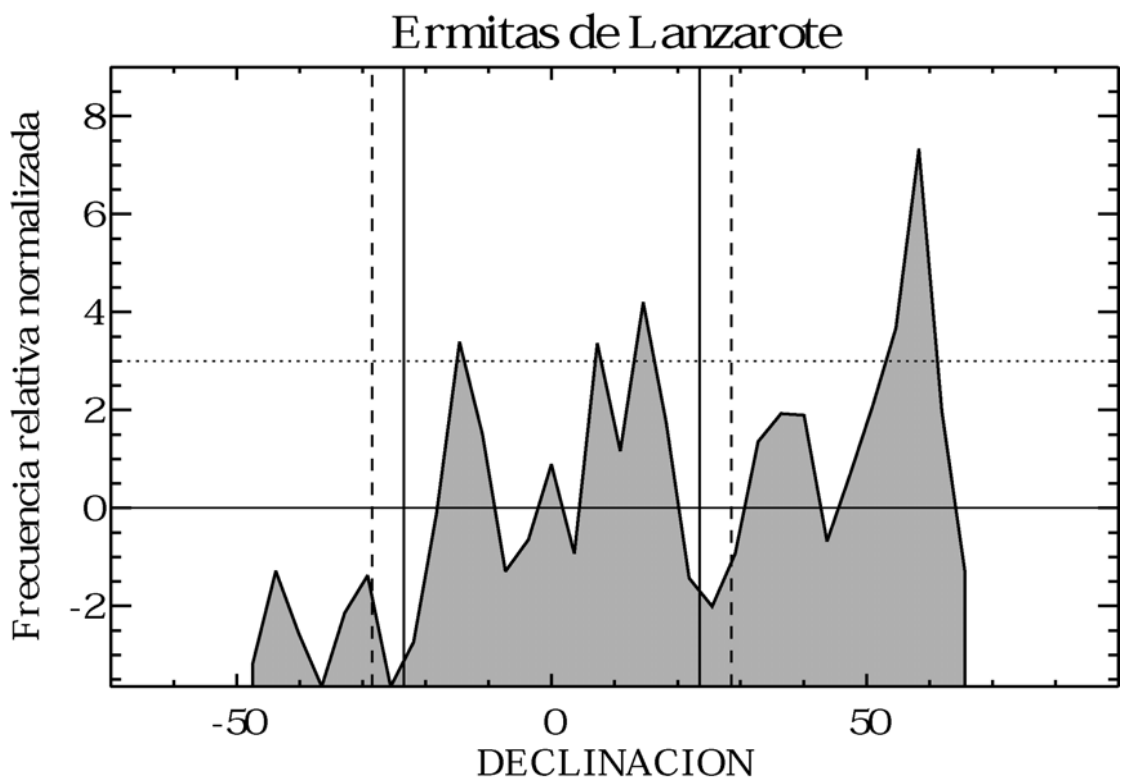

*Figura 7. Histograma de declinación para las ermitas e iglesias de Lanzarote obtenido de los datos de junio de 2012. Se encuentran unos pocos picos estadísticamente significativos, por encima del nivel 3σ. Las líneas verticales continuas representan las declinaciones correspondientes a las posiciones extremas del sol en los solsticios, mientras que las verticales discontinuas representan lo propio para la luna en los lunasticios mayores. Tres picos menores estadísticamente significativos se hallan en el rango solar (orientación canónica). Sin embargo, el pico más elevado, ubicado en torno a 58º es excepcional y podría asociarse a un pico de acumulación por orientaciones cercanas a la línea meridiana. Ver el texto para más detalles.*

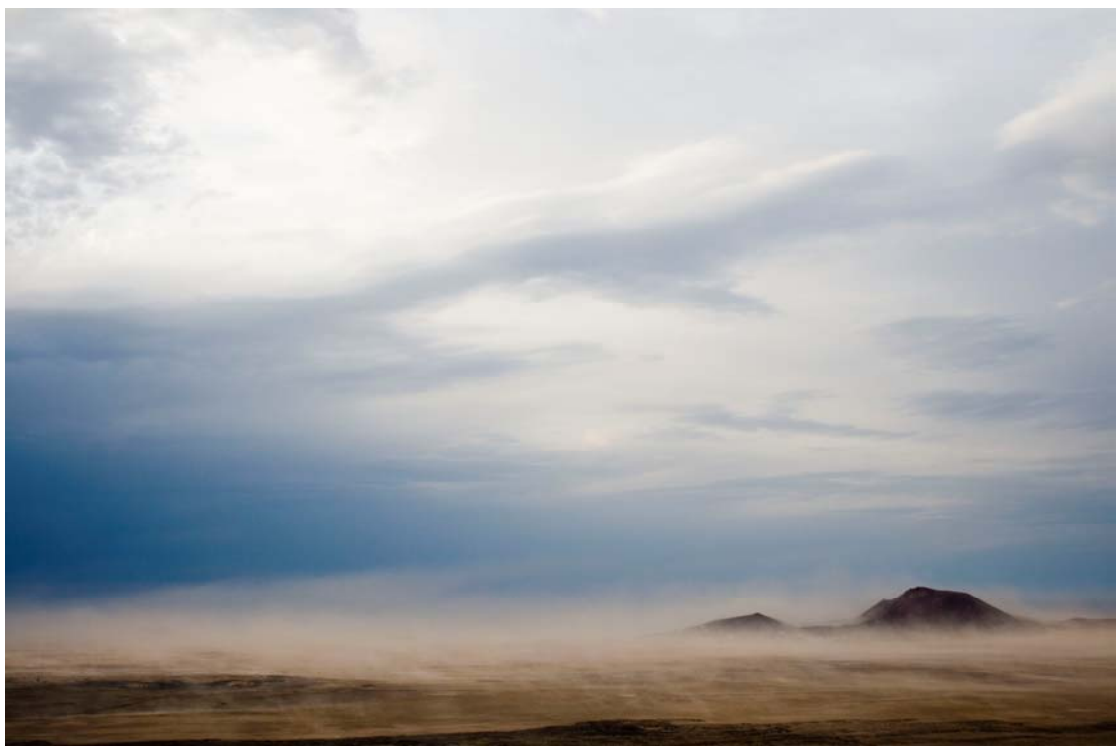

*Figura 8.* Tormenta reciente de arena en el Jable. Durante siglos, el flujo de las arenas ha alterado y modificado el paisaje de Lanzarote haciendo con frecuencia difícil la convivencia en los márgenes de esta comarca. Imagen, cortesía de José Farray.

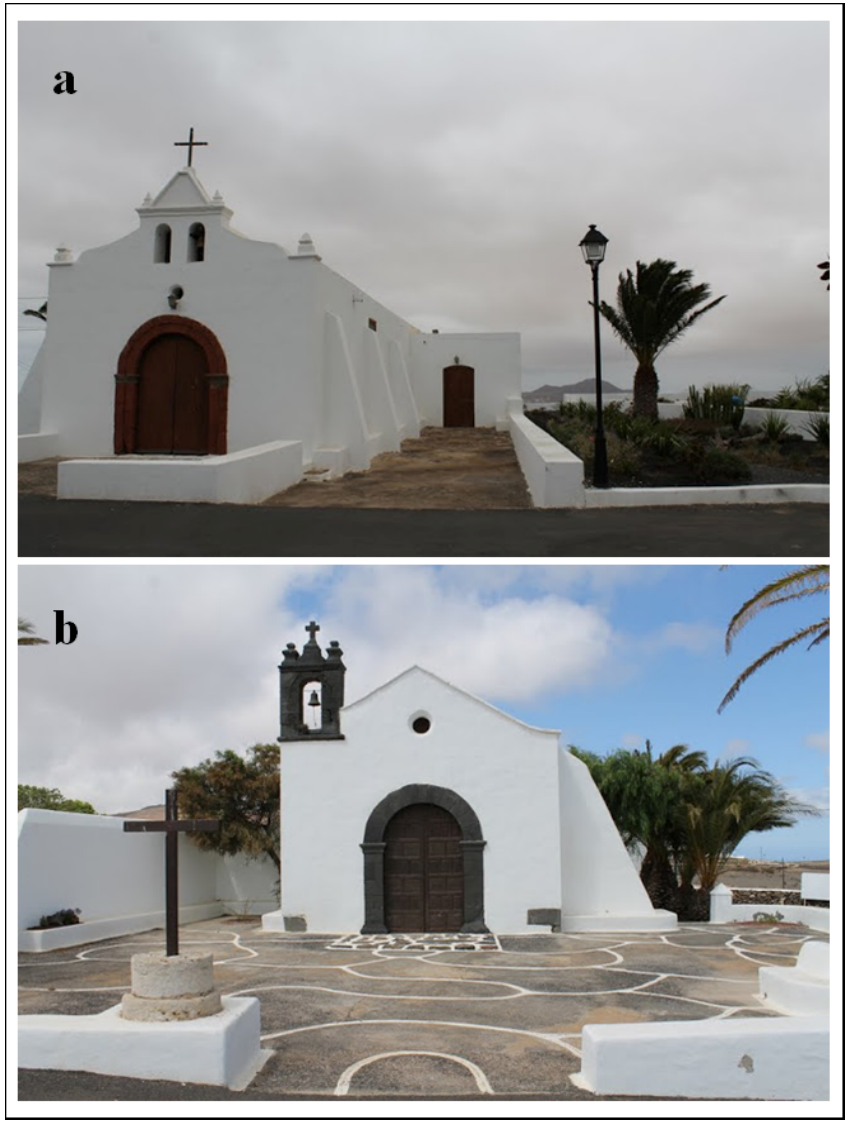

*Figura 9. Dos bellas ermitas, y bastante antiguas, orientadas en el rango de acimuts exclusivo de la isla de Lanzarote: (a) El Socorro de Tiagua, con un acimut de 12º enfrenta de lleno los vientos dominantes en la región del Jable. (b) San Sebastián del Mojón, orientada al NE, también cuenta con una barbacana cuya misión es la de proteger del viento el entorno de la portada de la construcción, aun expuesta. Imágenes de Alejandro Gangui.*

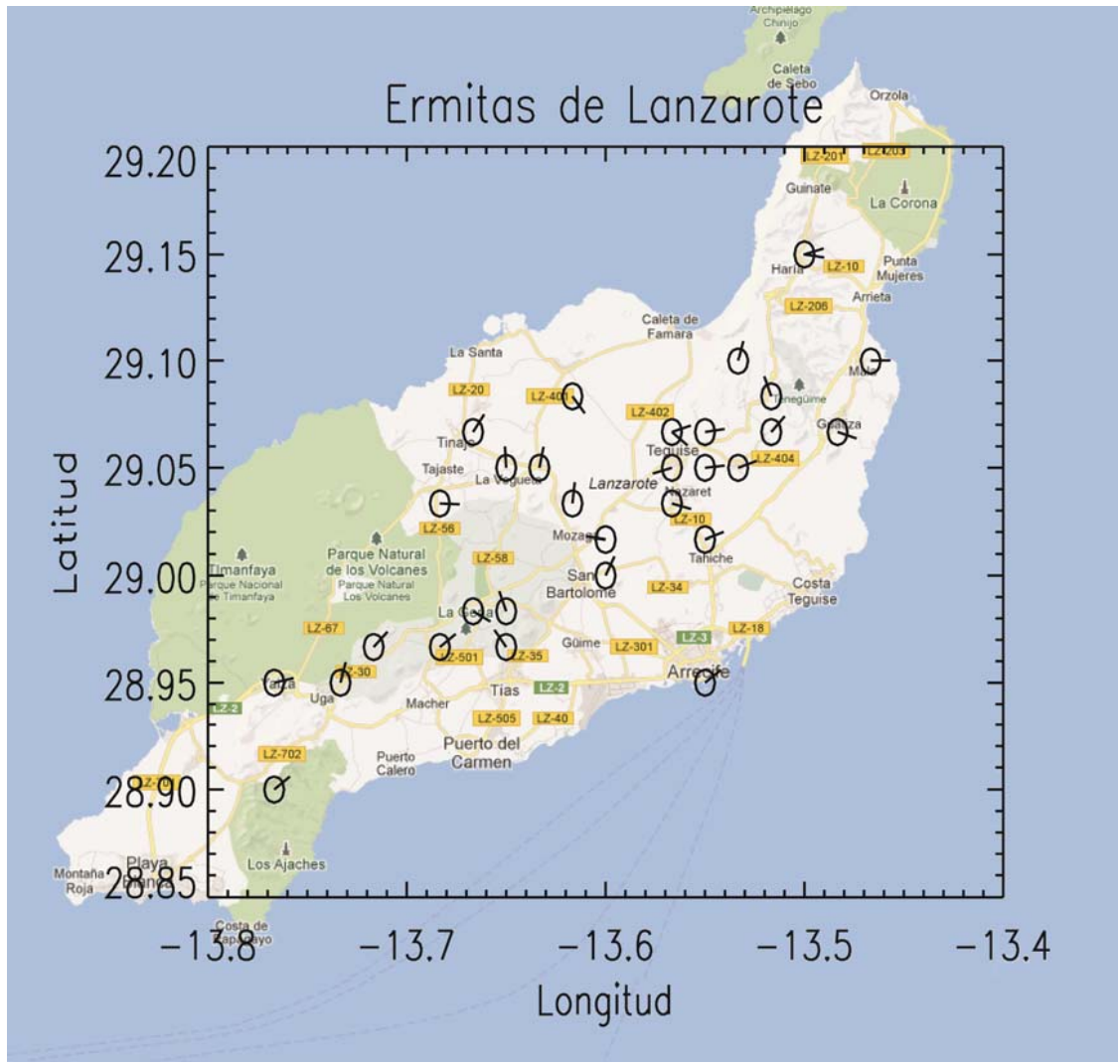

*Figura 10. Mapa con la ubicación geográfica de la totalidad de las iglesias medidas (señaladas con elipses), junto con la orientación del eje de las construcciones en dirección al ábside (rayas, orientadas de acuerdo a los acimuts consignados en la Tabla 1). En la ciudad de Haría, dos iglesias geográficamente muy próximas poseen orientaciones diferentes, lo que explica la presencia allí de una única elipse con dos rayas. Lo mismo sucede para un par de iglesias de la ciudad de Teguise. Imagen de los autores sobre un mapa cortesía de Google Maps.*

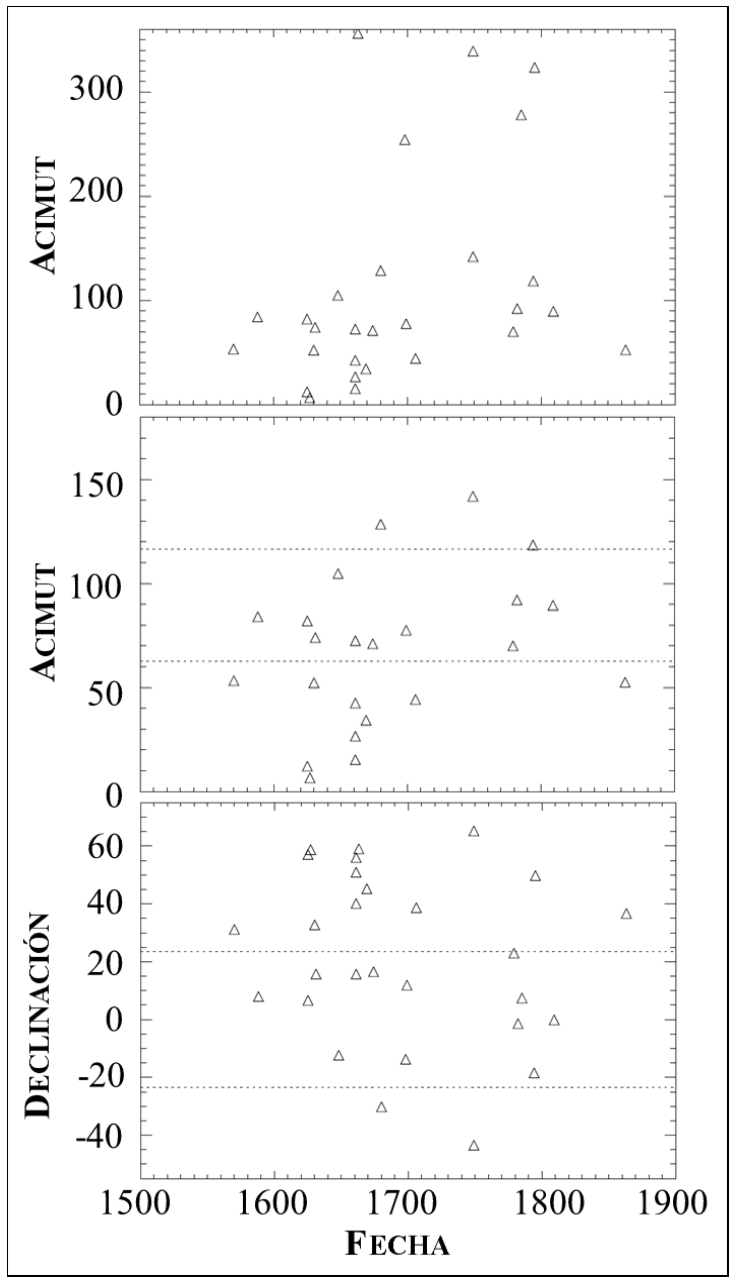

*Figura 11. (a) Diagrama que muestra el acimut de las iglesias frente a la fecha, donde se emplean las fechas más probables de construcción (o de su primera mención en las fuentes) de 28 de las 32 iglesias y ermitas estudiadas. (b) Igual que el diagrama anterior, pero donde se amplía la zona de acimuts cercana al rango solar. Excepto aquellas construcciones con clara orientación hacia el N o NE, una buena proporción de iglesias caen dentro del rango solar indicado por las dos líneas paralelas. (c) Diagrama de la declinación frente a la fecha. Las líneas horizontales señalan las declinaciones extremas donde se ubica el sol en los solsticios. Nótese el hueco en la construcción de ermitas asociado posiblemente a las erupciones de Timanfaya en la primera mitad del siglo XVIII.*